\documentclass[nobalancelastpage,aps,tightenlines,superscriptaddress,twocolumn,amssymb,nofootinbib]{revtex4-1}
\usepackage{amsmath, amsthm, amssymb} 

\usepackage{graphics,bm}
\usepackage{epsfig}
\usepackage{graphicx}
\usepackage{amsmath}
\usepackage{wrapfig}
\usepackage{color}

\newcommand{\beq}{\begin{equation}}
\newcommand{\eeq}{\end{equation}}
\newcommand{\bqa}{\begin{eqnarray}}
\newcommand{\eqa}{\end{eqnarray}}

\newcommand{\os}{\text{\tiny OS}}
\newcommand{\ms}{\overline{\text{\tiny MS}}}

\def\sumint{\hbox{$\sum$}\!\!\!\!\!\!\int}
\def\square{\vcenter{\vbox{\hrule height.4pt
          \hbox{\vrule width.4pt height4pt
          \kern4pt\vrule width.3pt}\hrule height.4pt}}}

\begin{document}

\title{Inhomogeneous chiral condensate in the quark-meson model}
\author{Prabal Adhikari}
\email{adhika1@stolaf.edu}
\affiliation{St. Olaf College, Physics Department, 1520 St. Olaf Avenue,
Northfield, Minnesota 55057, USA}

\author{Jens O. Andersen}
\email{andersen@tf.phys.ntnu.no}
\affiliation{Department of Physics, Faculty of Natural Sciences,NTNU, 
Norwegian University of Science and Technology, H{\o}gskoleringen 5,
N-7491 Trondheim, Norway}
\author{Patrick Kneschke}
\email{patrick.kneschke@uis.no}
\affiliation{Faculty of Science and Technology, University of Stavanger,
N-4036 Stavanger, Norway}
\date{\today}

\begin{abstract}
The two-flavor quark-meson model is used as a low-energy effective model for
QCD to study inhomogeneous chiral
condensates at finite 
baryon chemical potential $\mu_B$. 
The parameters of the model are determined by matching the meson and quark
masses, and the pion decay constant to their physical values using the on-shell and 
modified minimal subtraction schemes. 
Using a chiral-density wave ansatz for the
inhomogeneity, we calculate the effective potential in the mean-field
approximation and the result is completely analytic.
The size of the inhomogeneous phase depends 
sensitively on the pion mass and whether one includes the
vacuum fluctuations or not.
Finally, we briefly discuss the mean-field phase diagram.

\end{abstract}
\keywords{Dense QCD,
chiral transition, }
\maketitle
\hspace{-.45cm}
\section{Introduction}
The phase structure of QCD has been subject of interest since 
its phase diagram was first conjectured in the 1970s. Today, we have
a relatively good understanding of the phase transition at zero baryon
chemical potential $\mu_B$. At $\mu_B=0$ there is no sign problem and one can use lattice simulations. For 2+1 flavors and physical quark masses, the transition is 
a crossover at a temperature of around 155 MeV 
\cite{aoki1,aoki2,borsa,baza}. Above the transition temperature
QCD is in the quark-gluon plasma phase.
At temperatures up to a few times the transition temperature, 
this is a strongly interacting liquid \cite{kris}. For higher temperatures, 
resummed perturbation theory yields results 
for the thermodynamic functions that are in good agreement with lattice
data \cite{hon1,hon2}.

The situation is less clear at finite density and low temperature.
Due to the sign problem, this part of the phase diagram is not accessible
to standard Monte Carlo techniques based on importance sampling.
Only at asymptotically high densities are we confident about the
phase and the properties of QCD. 
In this limit, the ground state of QCD is the color-flavor locked phase
which is a color-superconducting phase \cite{rev1}.
The color symmetry is completely broken and all the gluons are screened.
The low-energy excitations of this phase are Goldstone modes which can
be described by a chiral effective Lagrangian.
At medium densities, information about the phase diagram has been obtained
mainly by using low-energy effective models that share some features with QCD
such as chiral symmetry breaking in the vacuum.
Examples of low-energy models are the Nambu-Jona-Lasinio (NJL)
model and the quark-meson (QM) model as well as their
Polyakov-loop extended versions PNJL and PQM models. 

Details and further motivation of the QM model can be found in \cite{zuber} and \cite{GellMann}, although historically the fermionic degrees of freedom were nucleons instead of quarks. One may object to having both quark and mesonic degrees of freedom present at the same time in the QM model, since quarks are confined at low temperatures. The Polyakov loop is introduced in order to mimic confinement in QCD in a statistical sense by coupling the chiral models to a constant $SU(N_c)$ background gauge field $A_{\mu}^a$ \cite{fukushima}, which is expressed in terms of the complex-valued Polyakov loop variable $\Phi$. Consequently
the effective potential becomes a function of the expectation value of the chiral condensate and the expectation value of the Polykov loop, where the latter then serves
as an approximate order parameter for confinement. Finally, one adds the contribution to the free energy density from the gluons via a phenomenological Polyakov loop
potential \cite{fukushima}.

At these lower densities, QCD is still in
a color-superconducting phase, but the
symmetry-breaking pattern is different \cite{rev1,rev2}.
The ground state for a given 
value of the baryon chemical potential is very sensitive to the
values of the parameters of the effective models. 
It turns out that some of the color-superconducting phases are
inhomogeneous \cite{rev1,rev2,suprev}.
Inhomogeneous phases do not exist only in dense QCD, but
also for example in ordinary superconductors and in imbalanced
Fermi gases.
In the present paper, we reconsider the problem of inhomogeneous
chiral-symmetry breaking phases in dense QCD \cite{buballarev,robd}
within the QM model. 
To be specific, we focus
on a chiral-density wave (CDW). 
The problem of inhomogeneous phases has been addressed before
in the context of the Ginzburg-Landau approach 
\cite{nick2,abuki,friman,japs}, the
NJL \cite{nakano,mada,nickel,vari,dirk,lee2} and PNJL 
models \cite{nickel2,brun}, the
QM model \cite{nickel,bubsc,fix1}, and the 
nonlocal chiral quark model \cite{scoc}.
Numerical methods for the calculation of the phase diagram for a general inhomogeneous condensate are available \cite{forcrand,wagner}, but we resort to a chiral-density wave ansatz in order to present analytical results. 

Most of the work has been done in the mean-field approximation; 
however, the properties of the Goldstone modes that are associated
with the spontaneous symmetry breaking of space-time symmetries are 
important as they may destabilize the inhomogeneous phase \cite{friman,japs}.
The destabilization is caused by long-wavelength fluctuations
at finite temperature, where long-range
order is replaced by algebraic decay of the order parameter.
This does not apply at $T=0$ since the long-wavelength fluctuations
are suppressed in this case. 

In the next section, we briefly discuss the QM model and 
explain how we calculate the one-loop
effective potential in the large-$N_c$ limit using the 
on-shell (OS)
and modified minimal subtraction
($\overline{\rm MS}$) schemes together with 
dimensional regularization. We also calculate analytically the
medium-dependent part of the effective potential and the
quark density at zero temperature.
In Sec. III, we present and discuss
our results for the different phases.
We also discuss the mean-field phase diagram as a function
of $T$ and $\mu$.
In Appendix \ref{integrali} we calculate some integrals and sum-integrals
that we need, and in Appendix \ref{appfix}, we calculate the 
parameters of the Lagrangian as functions of physical observables
to leading order in the large-$N_c$ expansion.
Finally, in Appendix \ref{epder},
we calculate the effective potential to the same order.

\section{Quark-meson model and effective potential}
The Euclidean Lagrangian of the two-flavor quark-meson model is 
\bqa\nonumber
{\cal L}&=&
{1\over2}\left[(\partial_{\mu}\sigma)^2
+(\partial_{\mu}{\boldsymbol \pi})^2\right]
+{1\over2}m^2(\sigma^2+{\boldsymbol\pi}^2)
\\&&
\nonumber
+{\lambda\over24}(\sigma^2+{\boldsymbol\pi}^2)^2
-h\sigma
\\&&
+\bar{\psi}_f
\left[
/\!\!\!\partial
-\gamma^0\mu_f
+g(\sigma+i\gamma^5{\boldsymbol\tau}\cdot{\boldsymbol\pi})\right]\psi_f\;,
\label{lag}
\eqa
where $f=u,d$ is the flavor index and $\mu_f$ is the corresponding chemical potential.
For $\mu_u=\mu_d$, in addition to a global
$SU(N_c)$ symmetry, the Lagrangian has a $U(1)_B\times SU(2)_L\times SU(2)_R$
symmetry in the chiral limit, while away from it, the symmetry is reduced
to $U(1)_B\times SU(2)_V$.
For $\mu_u\neq\mu_d$, the symmetry is reduced to
$U(1)_B\times U(1)_{I_3L}\times U(1)_{I_3R}$ for $h=0$
and $U(1)_B\times U(1)_{I_3}$ for $h\neq0$.
In the remainder of this paper we choose
$\mu_u=\mu_d=\mu={1\over3}\mu_B$, where 
$\mu$ is the quark chemical potential and $\mu_B$ is the baryon chemical 
potential. 

In the vacuum, the $\sigma$ field acquires a nonzero vacuum expectation
value, which we denote by $\phi_0$. 
We next make an ansatz for the inhomogeneity.
In the literature, mainly one-dimensional modulations have been considered,
for example CDW and soliton lattices.
Since the results seem fairly independent of the 
modulation \cite{bubsc}, we opt for the simplest, namely 
a one-dimensional chiral-density wave. The ansatz is
\bqa
\sigma(z)=
\phi_0\cos(qz)
\;,
\hspace{0.5cm}
\pi_3(z)=
\phi_0\sin(qz)\;,
\label{back}
\eqa
where $\phi_0$ is the magnitude of the wave and $q$ is a wave vector.
The mean fields can be combined into a complex order parameter
$M(z)=g[\sigma(z)+i\pi_3(z)]=\Delta e^{iqz}$,
where $\Delta=g\phi_0$. The dispersion relation of the quarks in the 
background (\ref{back}) is known \cite{dautry}
\bqa
E_{\pm}^2&=&\left(\sqrt{p_{\parallel}^2+\Delta^2}\pm{q\over2}\right)^2+p_{\perp}^2\;,
\label{disp}
\eqa
where $p_{\parallel}=p_3$ and $p_{\perp}^2=p_1^2+p_2^2$.
In the QCD vacuum, the chiral symmetry is broken by forming
pairs of left-handed quarks  and right-handed antiquarks (and vice versa).
These quark-antiquark pairs have zero net momentum and so the
chiral condensate is homogeneous with $q=0$. An inhomogeneous chiral condensate
in the vacuum would imply the spontaneous breakdown of rotational symmetry.
At finite density, it is possible to form an inhomogeneous condensate
by pairing a left-handed quark with a right-handed quark with the same
momentum. The net momentum of the pair is nonzero, resulting in an
inhomogeneous chiral condensate.
A nonzero wave vector $q$ lowers the energy of the negative branch
in (\ref{disp}) and as a result only this branch is occupied by the
quarks in this phase \cite{buballarev}.

At tree level, the parameters of the Lagrangian (\ref{lag})
$m^2$, $\lambda$, $g^2$, and $h$
are related to the the physical quantities 
$m_\sigma^2$, $m_\pi^2$, $m_q$, and $f_\pi$ by
\bqa
\label{tr1}
&&m^2=-{1\over2}\left(m_{\sigma}^2-3m_{\pi}^2\right)\;,\;
\lambda=3{(m_{\sigma}^2-m_{\pi}^2)\over f_{\pi}^2}\;,\;
\\ &&
g^2={m_q^2\over f_{\pi}^2}
\;,\;\;
h=m_\pi^2f_\pi
\;.
\label{tr4}
\eqa
Expressed in terms of physical quantities, the tree-level potential is
\bqa\nonumber
V_{\rm tree}&=&
{1\over2}f_\pi^2q^2{\Delta^2\over m_q^2}
-{1\over4}f_\pi^2(m_{\sigma}^2-3m_{\pi}^2){\Delta^2\over m_q^2}
\\ &&
+{1\over8}f_\pi^2(m_{\sigma}^2-m_{\pi}^2){\Delta^4\over m_q^4}
-m_\pi^2f_\pi^2{\Delta\over m_q}\;.
\label{tri}
\eqa
The relations in Eqs. (\ref{tr1})--(\ref{tr4}) 
are the parameters determined
at tree level and are often used in practical calculations. However, 
this is inconsistent in calculations
that involve loop corrections unless one uses the OS
renormalization scheme.
In the on-shell scheme, the divergent loop integrals are regularized
using dimensional regularization, but the counterterms are chosen 
differently from the 
($\overline{\rm MS}$)
scheme. 
The counterterms in the on-shell 
scheme are chosen so that they
exactly cancel the loop corrections to the self-energies and couplings
evaluated on shell, and as a result the
renormalized parameters are independent of the renormalization scale
and satisfy the tree-level relations (\ref{tr1})--(\ref{tr4}).
In the $\overline{\rm MS}$ scheme,  
the relations (\ref{tr1})--(\ref{tr4}) receive radiative corrections
and the parameters depend on the renormalization scale.
The divergent part of a counterterm in the OS scheme is necessarily
the same as the counterterm in the $\overline{\rm MS}$ scheme.
Since the bare parameters are independent of the renormalization scheme,
one can write down relations between the renormalized parameters
in the $\overline{\rm MS}$ and the OS scheme.
The latter are expressed in terms of the physical masses and couplings
in Eqs. (\ref{tr1})--(\ref{tr4}) and we can therefore express the 
renormalized running parameters 
$m_{\ms}^2$, $\lambda_{\ms}$, $g_{\ms}^2$, and $h_{\ms}$
in the $\overline{\rm MS}$ scheme
in terms of the masses
$m_{\sigma}^2$, $m_\pi^2$, and $m_q$, and the pion decay constant $f_\pi$.
In Ref. \cite{onshell}, we calculated the parameters 
in the chiral limit. 
In this paper we gene\-ralize these relations
to the physical point, which are derived in Appendix \ref{appfix}.
The result for the renormalized
one-loop effective potential in the large-$N_c$
limit is derived in Appendix \ref{epder} and reads
\begin{widetext}
\bqa\nonumber
V_{\rm1-loop}&=&{1\over2}f_{\pi}^2q^2
\left\{1-\dfrac{4 m_q^2N_c}{(4\pi)^2f_\pi^2}
\left[\log\mbox{$\Delta^2\over m_q^2$}
+F(m_\pi^2)+m_\pi^2F^{\prime}(m_\pi^2)\right]
\right\}{\Delta^2\over m_q^2}
+\dfrac{3}{4}m_\pi^2 f_\pi^2
\left\{1-\dfrac{4 m_q^2N_c}{(4\pi)^2f_\pi^2}m_\pi^2F^{\prime}(m_\pi^2)
\right\}\dfrac{\Delta^2}{m_q^2}
\\ \nonumber &&
 -\dfrac{1}{4}m_\sigma^2 f_\pi^2
\left\{
1 +\dfrac{4 m_q^2N_c}{(4\pi)^2f_\pi^2}
\left[ \left(1-\mbox{$4m_q^2\over m_\sigma^2$}
\right)F(m_\sigma^2)
 +\dfrac{4m_q^2}{m_\sigma^2}
-F(m_\pi^2)-m_\pi^2F^{\prime}(m_\pi^2)
\right]\right\}\dfrac{\Delta^2}{m_q^2} \\ \nonumber
 & & + \dfrac{1}{8}m_\sigma^2 f_\pi^2
\left\{ 1 -\dfrac{4 m_q^2  N_c}{(4\pi)^2f_\pi^2}\left[
\dfrac{4m_q^2}{m_\sigma^2}
\left( 
\log\mbox{$\Delta^2\over m_q^2$}
-\mbox{$3\over2$}
\right) -\left( 1 -\mbox{$4m_q^2\over m_\sigma^2$}\right)F(m_\sigma^2)
+F(m_\pi^2)+m_\pi^2F^{\prime}(m_\pi^2)\right]
 \right\}\dfrac{\Delta^4}{m_q^4}
\\ \nonumber&&
- \dfrac{1}{8}m_\pi^2 f_\pi^2
\left[1-\dfrac{4 m_q^2N_c}{(4\pi)^2f_\pi^2}m_\pi^2F^{\prime}(m_\pi^2)\right]
\dfrac{\Delta^4}{m_q^4}
-m_\pi^2f_\pi^2\left[
1-\dfrac{4 m_q^2  N_c}{(4\pi)^2f_\pi^2}m_\pi^2F^{\prime}(m_\pi^2)
\right]\dfrac{\Delta}{m_q}
-{N_cq^4\over6(4\pi)^2}
\\ \nonumber&&
+{N_c\over3(4\pi)^2}
\left[q\sqrt{{q^2\over4}-\Delta^2}(26\Delta^2+q^2)
-12\Delta^2(\Delta^2+q^2)\log{{q\over 2}+\sqrt{{q^2\over4}-\Delta^2}\over\Delta}
\right]
\theta(\mbox{$q\over2$}-\Delta)
\\&&
-2N_cT\int_p\bigg\{
\log\left[1+e^{-\beta(E_{\pm}-\mu)}\right]
+\log\left[1+e^{-\beta(E_{\pm}+\mu)}\right]\bigg\}\;,
\label{fullb}
\eqa
\end{widetext}
where $E_{\pm}$ is given by Eq. (\ref{disp})
and a sum over $\pm$ is implied.
Moreover, $F(p^2)$ and $F^{\prime}(p^2)$ are defined in Appendix \ref{integrali}.
We note that the vacuum part of the effective potential
(obtained by setting $q=\mu=T=0$) of Eq. (\ref{fullb})
has its minimum at $\Delta=m_q$ by construction, as does the tree-level
potential Eq. (\ref{tri}).
The result for the vacu\-um part of the effective potential is completely
analytic and obtained using dimensional regularization.
At this point, a few remarks on the regularization of the
effective potential are appropriate. 
A physically meaningful effective
potential cannot depend on the wave vector $q$ when the amplitude
$\Delta$ vanishes. It is straightforward to show that the
$T=\mu=0$ part of Eq. (\ref{fullb}) satisfies this.
The finite $T/\mu$ part of the effective potential, i.e. 
the last line of Eq. (\ref{fullb}) also satisfies this, but at finite $T$ one must show it numerically. At $T=0$, it can be 
show analytically, see below.
If one regularizes the effective potential
with a sharp momentum cutoff $\Lambda$ \cite{consistent}, it is not
independent of $q$ for $\Delta=0$
The residual $q$ dependence in the
limit $\Delta\rightarrow0$ is then an artifact of the regulator which can be 
dealt with by introducing extra subtraction terms. Different regularization
methods are discussed in some
detail in \cite{symren1,consistent}.

In the limit $T=0$, we can calculate the medium contribution 
to the effective potential $V_{\rm 1-loop}$ analytically.
Since this contribution is finite, the calculation can be done directly
in three dimensions. This contribution
is given by the zero-temperature limit of the last line
in Eq. (\ref{fullb}) and is denoted by $V_1^{\rm med}$. 
We first consider the contribution from $E_+$
in Eq. (\ref{fullb}), which we denote by $V_{1+}^{\rm med}$. 
At $T=0$, this reads
\bqa\nonumber
V_{1+}^{\rm med}&=&-2N_c\int_p(\mu-E_+)\theta(\mu-E_+)\\
\nonumber
&=&-{16N_c\over(4\pi)^2}
\int_0^{\infty}dp_{\parallel}
\\ &&
\times\int_0^{\infty}
(\mu-E_+)\theta(\mu-E_+)p_{\perp}dp_{\perp}\;.
\eqa
The integral over $p_{\perp}$ is straightforward
to do, but we have to be careful
with the upper limit due to the step function.
The upper limit, denoted by $p_{\perp}^f$, is a function of $p_{\parallel}$
and is given by
\bqa
(p_{\perp}^f)^2&=&\mu^2-\left(\sqrt{p_{\parallel}^2+\Delta^2}+{q\over2}\right)^2\;.
\label{pperpp}
\eqa
\begin{widetext}
Integrating over $p_{\perp}$ from $p_{\perp}=0$ to $p_{\perp}=p_{\perp}^f$
yields
\bqa
V_{1+}^{\rm med} 
&=&-{16N_c\over(4\pi)^2}\int_0^{p_{\parallel}^f}
\left[
{1\over6}\mu^3+{1\over3}\left(\sqrt{p_{\parallel}^2+\Delta^2}+{q\over2}\right)^3
-{1\over2}\mu\left(\sqrt{p_{\parallel}^2+\Delta^2}+{q\over2}\right)^2
\right]dp_{\parallel}\;,
\eqa
where the upper limit of integration is denoted by $p_{\parallel}^f$. 
The upper limit
can be found by setting $p_{\perp}=0$ in the dispersion relation
or $p_{\perp}^f=0$ in (\ref{pperpp}) 
and is therefore given by
\bqa
p_{\parallel}^f&=&\sqrt{\left(\mu - {q\over2}\right)^2 - \Delta^2}\;.
\eqa
Changing variables to $u=\sqrt{p_{\parallel}^2+\Delta^2}$, we obtain
\bqa
V_{1+}^{\rm med}
&=&-{16N_c\over(4\pi)^2}
\int_{\Delta}^{u_{+}^f}
\left[
{1\over6}\mu^3+{1\over3}\left(u+{q\over2}\right)^3
-{1\over2}\mu\left(u+{q\over2}\right)^2
\right] {u\;du\over\sqrt{u^2-\Delta^2}}\;,
\label{a1u}
\eqa
where the upper limit is $u_{+}^f=\mu-{q\over2}$.
In order to get a nonzero contribution, we must have $\mu\geq\Delta+{q\over2}$.
Integrating over $u$, we find
\bqa\nonumber
V_{1+}^{\rm med}
&=&-{2N_c\over(4\pi)^2}
\left[{2\over3}
\sqrt{\left(\mu-{q\over2}\right)^2-\Delta^2}
\left[
\left(\mu+{q\over2}\right)\left(\mu-{q\over2}\right)^2
+{1\over4}\Delta^2(13q-10\mu)\right]
\right.\\ &&\left.
+\Delta^2(\Delta^2-2\mu q+q^2)
\log{\mu-{q\over2}+\sqrt{(\mu-{q\over2})^2-\Delta^2}\over\Delta}
\right]\theta(\mu-\mbox{$q\over2$}-\Delta)
\;.
\label{cont1}
\eqa
The second contribution is for $E_-$ in Eq. (\ref{fullb}). It is 
denoted by $V_{1-}^{\rm med}$ and is found from Eq. (\ref{a1u})
by the substitution $q\rightarrow -q$. This gives
\bqa
V_{1-}^{\rm med}
&=&-{16N_c\over(4\pi)^2}
\int_{u_{\rm low}}^{u_{-}^f}
\left[
{1\over6}\mu^3+{1\over3}\left|u-{q\over2}\right|^3
-{1\over2}\mu\left(u-{q\over2}\right)^2
\right]{u\;du\over\sqrt{u^2-\Delta^2}}\;,
\eqa
where the upper limit is $u_{-}^f=\mu+{q\over2}$ and the lower limit
is $u_{\rm low}$. The lower limit depends on the relative magnitude of 
$\mu$, $\Delta$, and ${q\over2}$. The different cases are discussed below.
\begin{enumerate}
\item{$\Delta>{q\over2}$: The dispersion relation is shown in the left
panel of Fig. \ref{dispi}.
$u_{\rm low}=\Delta$}
and there is a nonzero contribution if
$\mu-\Delta+{q\over2}>0$. This contribution is obtained from (\ref{cont1})
by the substitution $q\rightarrow -q$. This yields
\bqa\nonumber
V_{1-}^{\rm med}
&=&-{2N_c\over(4\pi)^2}
\left[
{2\over3}\sqrt{\left(\mu+{q\over2}\right)^2-\Delta^2}
\left[\left(\mu-{q\over2}\right)\left(\mu+{q\over2}\right)^2
-{1\over4}\Delta^2(13q+10\mu)
\right]
\right.\\ &&\left.
+\Delta^2(\Delta^2+2\mu q+q^2)
\log{\mu+{q\over2}+\sqrt{(\mu+{q\over2})^2-\Delta^2}\over\Delta}
\right]\theta(\mu+\mbox{$q\over2$}-\Delta)\;.
\label{med2}
\eqa
\item{$\Delta<{q\over2}$: The dispersion relation is shown in the right panel
of Fig. \ref{dispi} (blue curve) and 
the minimum of $|E_-|$
is at $p=p_0=\sqrt{{q^2\over4}-\Delta^2}$ and is zero.
For $p<p_0$, we have $|E_-|={q\over2}-u$ and for $p>p_0$, we 
have $|E_-|=u-{q\over2}$.}
For $\Delta<{q\over2}$, we also have to distinguish between the cases
$\mu>{q\over2}-\Delta$ and $\mu<{q\over2}-\Delta$.
\begin{enumerate}
\item{If $\mu>{q\over2}-\Delta$},
we have to integrate from $p=0$ to 
$p=p_f=\sqrt{(\mu+{q\over2})^2-\Delta^2}$, or from $u_{\rm low}=\Delta$
to $u=\mu+{q\over2}$. 
The green horizontal line indicates the
value of the chemical potential and the intersection with the dispersion
relation gives the upper limit of integration.
This yields
\bqa \nonumber
V_{1-}^{\rm med}&=&-{2N_c\over(4\pi)^2}
\left[
{2\over3}\sqrt{\left(\mu+{q\over2}\right)^2-\Delta^2}
\left[\left(\mu-{q\over2}\right)\left(\mu+{q\over2}\right)^2
-{1\over4}\Delta^2(13q+10\mu)
\right]
\right.\\ && \nonumber
\left.
+\Delta^2(\Delta^2+2\mu q+q^2)
\ln{\mu+{q\over2}+\sqrt{(\mu+{q\over2})^2-\Delta^2}\over\Delta}
\right.\\ &&\left.
+{1\over6}q\sqrt{{q^2\over4}-\Delta^2}(26\Delta^2+q^2)
-2\Delta^2(\Delta^2+q^2)\ln{{q\over2}+\sqrt{{q^2\over4}-\Delta^2}\over\Delta}
\right]\theta(\mu-\mbox{$q\over2$}+\Delta)
\;.
\label{cont3}
\eqa
\item{If $\mu<{q\over2}-\Delta$, we must integrate from
$p=\sqrt{(\mu-{q\over2})^2-\Delta^2}$ to
$p=p_f=\sqrt{(\mu+{q\over2})^2-\Delta^2}$, or from $u_{\rm low}={q\over2}-\mu$
to $u_{-}^f=\mu+{q\over2}$. 
The value of the chemical potential is
indicated by the orange line and the intersection with the dispersion relation
gives the upper and lower limits of integration.
This yields}
\bqa \nonumber
V_{1-}^{\rm med}&=&
-{2N_c\over(4\pi)^2}
\left[
-{2\over3}\sqrt{\left(\mu-{q\over2}\right)^2-\Delta^2}
\left[\left(\mu+{q\over2}\right)\left(\mu-{q\over2}\right)^2
+{1\over4}\Delta^2(13q-10\mu)
\right]
\right.\\ && \nonumber\left.
+{2\over3}\sqrt{\left(\mu+{q\over2}\right)^2-\Delta^2}
\left[\left(\mu-{q\over2}\right)\left(\mu+{q\over2}\right)^2
-{1\over4}\Delta^2(13q+10\mu)
\right]
\right.\\ && \nonumber\left.
+\Delta^2(\Delta^2-2\mu q+q^2)
\ln{{q\over2}-\mu+\sqrt{(\mu-{q\over2})^2-\Delta^2}\over\Delta}
+\Delta^2(\Delta^2+2\mu q+q^2)
\ln{\mu+{q\over2}+\sqrt{(\mu+{q\over2})^2-\Delta^2}\over\Delta}
\right.\\ && \left.
+{1\over6}q\sqrt{{q^2\over4}-\Delta^2}(26\Delta^2+q^2)
-2\Delta^2(\Delta^2+q^2)\ln{{q\over2}+\sqrt{{q^2\over4}-\Delta^2}\over\Delta}
\right]\theta(\mbox{$q\over2$}-\mu-\Delta)
\;. 
\label{cont4}
\eqa
\end{enumerate}
\end{enumerate}
The expression for $V_{1+}^{\rm med}$ and 
the different expressions for $V_{1-}^{\rm med}$ can be combined to give
our final result for the matter-dependent part of the Eq. (\ref{fullb})
\bqa\nonumber
V_{1+}^{\rm med+}+V_{1-}^{\rm med}
&=&-{2N_c\over(4\pi)^2}
\left\{{2\over3}
\sqrt{\left(\mu-{q\over2}\right)^2-\Delta^2}
\left[
\left(\mu+{q\over2}\right)\left(\mu-{q\over2}\right)^2
+{1\over4}\Delta^2(13q-10\mu)\right]{\rm sign}(\mu-\mbox{$q\over2$})
\right.\\ \nonumber
&&\left.
+\Delta^2(\Delta^2-2\mu q+q^2)
\log{|\mu-{q\over2}|+\sqrt{(\mu-{q\over2})^2-\Delta^2}\over\Delta}
\right\}\theta(|\mu-\mbox{$q\over2$}|-\Delta)
\\  \nonumber
&& 
-{2N_c\over(4\pi)^2}
\left\{
{2\over3}\sqrt{\left(\mu+{q\over2}\right)^2-\Delta^2}
\left[\left(\mu-{q\over2}\right)\left(\mu+{q\over2}\right)^2
-{1\over4}\Delta^2(13q+10\mu)
\right]
\right.\\ \nonumber
&&\left.
+\Delta^2(\Delta^2+2\mu q+q^2)
\log{\mu+{q\over2}+\sqrt{(\mu+{q\over2})^2-\Delta^2}\over\Delta}
\right\}\theta(\mu+\mbox{$q\over2$}-\Delta)
\\ &&
-{N_c\over3(4\pi)^2}
\left[q\sqrt{{q^2\over4}-\Delta^2}(26\Delta^2+q^2)
-12\Delta^2(\Delta^2+q^2)\log{{q\over 2}+\sqrt{{q^2\over4}-\Delta^2}\over\Delta}
\right]
\theta(\mbox{$q\over2$}-\Delta)
\label{totmed}
\;.
\eqa
Setting $\Delta=0$ in Eq. (\ref{totmed}), it is straightforward
to verify
that the matter part of the effective potential is independent
of $q$, as discussed after Eq. (\ref{fullb}).
Moreover, we note that the last line of Eq. (\ref{totmed}) cancels against
the penultimate line in Eq. (\ref{fullb}) in the complete thermodynamic
potential.

\begin{figure}
\begin{center}
\includegraphics[width=0.44\textwidth]{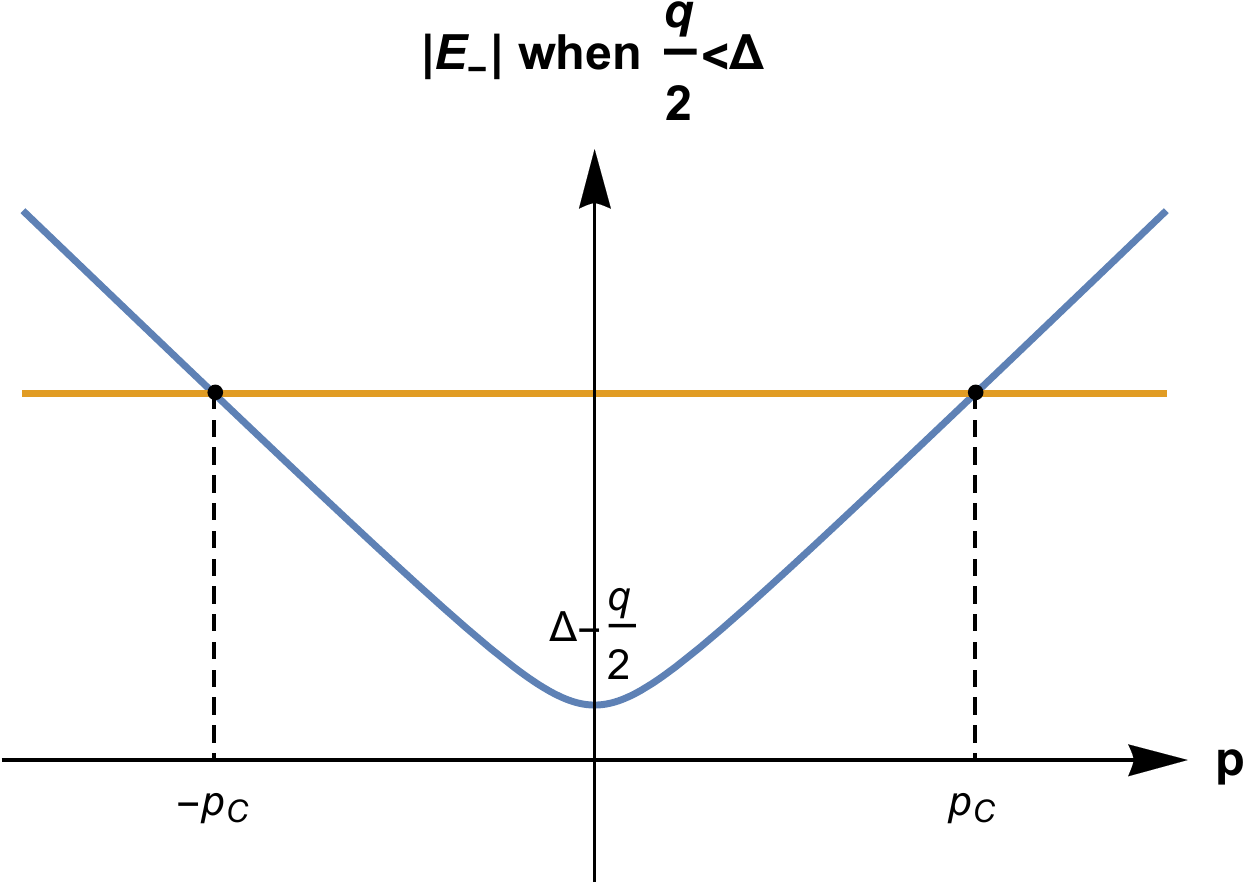}
\includegraphics[width=0.44\textwidth]{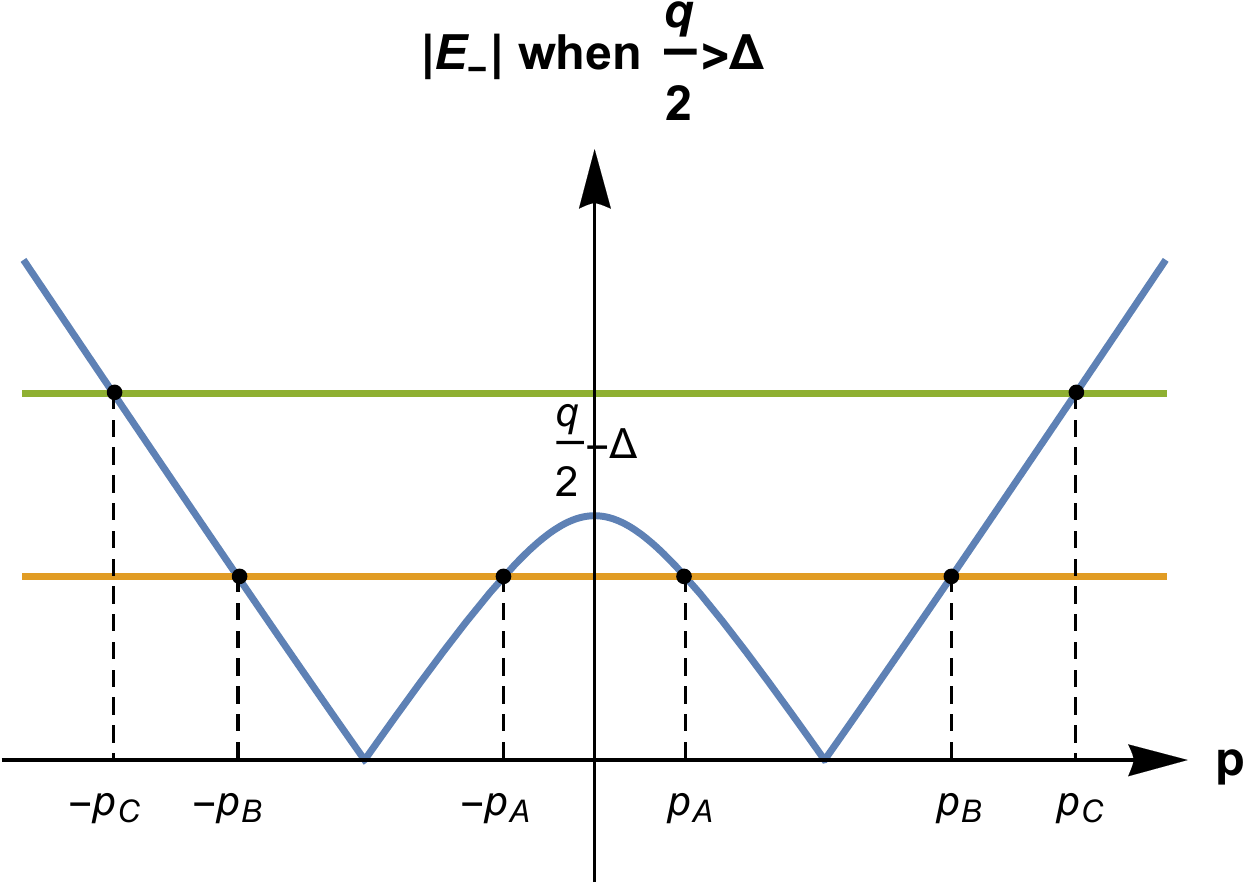}
\end{center}
\caption{Left: Dispersion relation $E_-$
for ${q\over2}<\Delta$. The horizontal orange line is for $\mu > \Delta-{q\over 2}$. Right: Dispersion relation $E_-$ for ${q\over2}<\Delta$.
The horizontal green line is for the case $\mu>{q\over2}-\Delta$
and the horizontal orange line is for the case $\mu<{q\over2}-\Delta$.
See main text for discussion of the regions of integration in the different
cases.}
\label{dispi}
\end{figure}
In the limit $T=0$, we can obtain an analytic result
for the quark density $n_q$ as well.
It is given by 
\bqa\nonumber
n_q&=&-{\partial V_{1+}^{\rm med}\over\partial\mu}
-{\partial V_{1-}^{\rm med}\over\partial\mu}
=
2N_c\int_p\left[\theta(\mu-E_+)+\theta(\mu-E_-)\right]
\\ \nonumber
&=&
{4N_c\over(4\pi)^2}
\left[{2\over3}
\sqrt{\left(\mu+{q\over2}\right)^2-\Delta^2}
\left(\mu^2-\Delta^2+{1\over4}\mu q-{q^2\over8}
\right)
+\Delta^2q
\log{\mu+{q\over2}+\sqrt{(\mu+{q\over2})^2-\Delta^2}\over\Delta}
\right]\theta(\mu+\mbox{$q\over2$}-\Delta)
\\ \nonumber &&
+{4N_c\over(4\pi)^2}
\left[{2\over3}
\sqrt{\left(\mu-{q\over2}\right)^2-\Delta^2}
\left(\mu^2-\Delta^2-{1\over4}\mu q-{q^2\over8}
\right){\rm sign}(\mu-\mbox{$q\over2$})
-\Delta^2q
\log{|\mu-{q\over2}|+\sqrt{(\mu-{q\over2})^2-\Delta^2}\over\Delta}
\right]
\\ &&
\times
\theta(|\mu-\mbox{$q\over2$}|-\Delta)
\;.
\label{qd}
\eqa
The quark density (\ref{qd}) is also independent of the wave vector $q$ when
the amplitude $\Delta$ is set to zero, as can be verified by inspection.

\end{widetext}
\section{Results and discussion}

In the numerical work, we set $N_c=3$ everywhere.
We use a constituent quark mass $m_q=300$ MeV. Since the sigma mass
is not very well known experimentally \cite{databook}, one 
typically allows it to vary between
$m_\sigma=400$ MeV and $m_\sigma=800$ MeV. 
We choose $m_\sigma=600$ MeV. 
At the physical point we take $m_\pi=140$ MeV and
for the pion decay constant we use $f_\pi=93$ MeV. In the the chiral limit the pion mass is zero.



\begin{figure}[htb]
\includegraphics[width=0.45\textwidth]{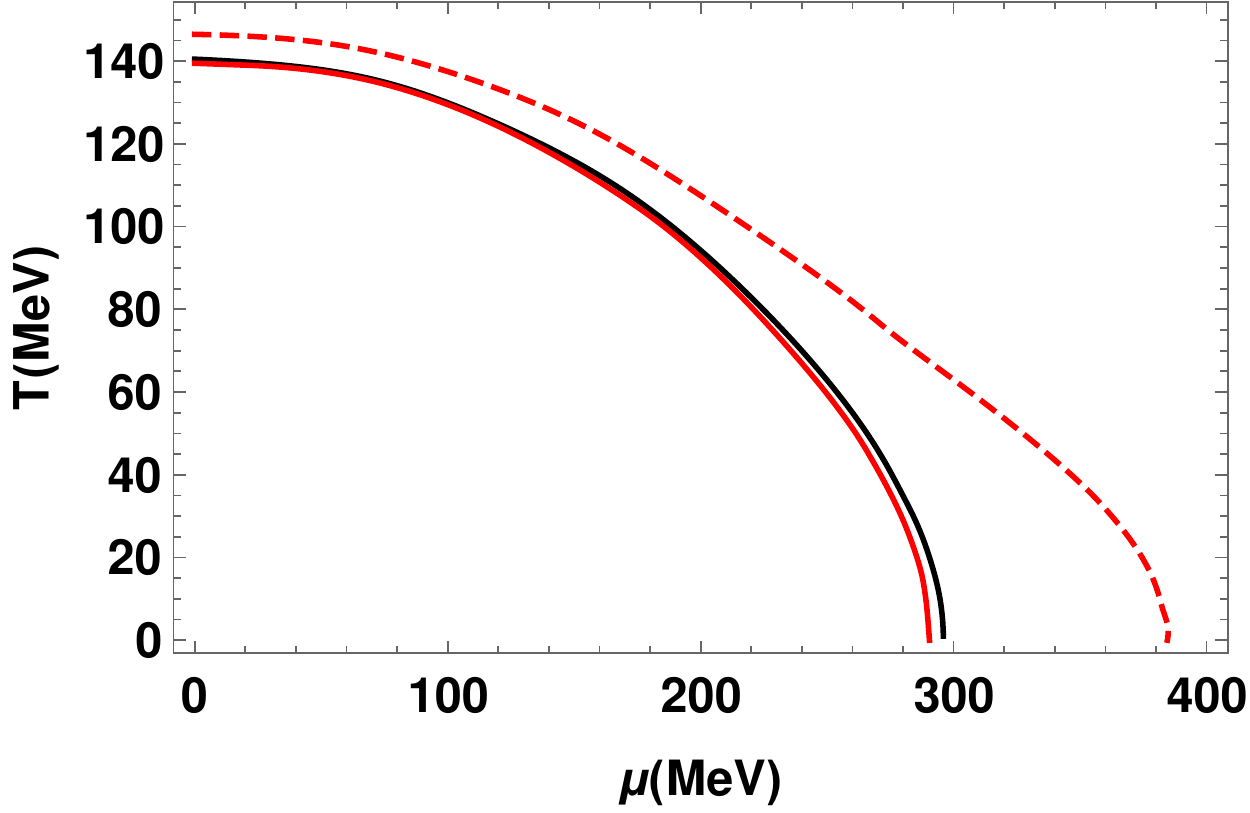}
\caption{The phase diagram in the $\mu$--$T$ plane 
for $m_q=300$ MeV and
$m_{\sigma}=600$ MeV in the chiral limit without quantum fluctuations.
A dashed line indicates a second-order transition, while a 
solid line indicates a first-order transition. 
The region between the red lines is the inhomogeneous phase.}
\label{chiral}
\end{figure}

It is known from earlier studies in the homogeneous case that
vacuum fluctuations play an important role. If we omit 
the quantum fluctuations, the phase transition in the chiral limit
is first order in the entire $\mu$--$T$ plane. If they are included
the transition is first order for $T=0$ and second order for $\mu=0$.
The first-order line starting on the $\mu$ axis ends at a tricritical
point. 
In the inhomogeneous case, we therefore examine the importance
of these fluctuations as well. 
In Fig. \ref{chiral}, we show the phase diagram in the $\mu$--$T$ plane 
in the chiral limit without vacuum fluctuations. The solid lines
indicate a first-order transition while the dashed line indicates
a second-order transition. The region between the two red lines is
the inhomogeneous phase. The black line is the first-order transition line
in the homogeneous case.

\begin{figure}[htb]
\includegraphics[width=0.45\textwidth]{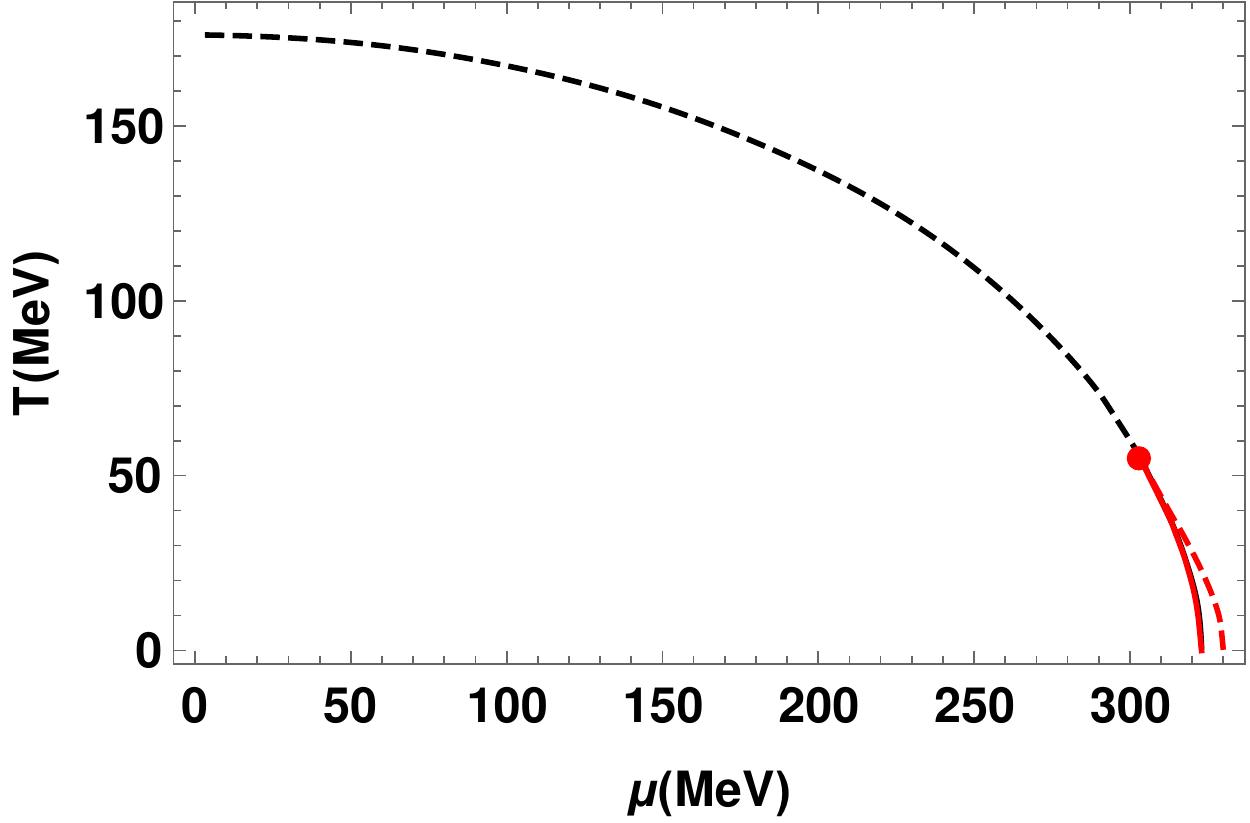}
\caption{The phase diagram in the $\mu$--$T$ plane 
for $m_q=300$ MeV and
$m_{\sigma}=600$ MeV in the chiral limit including vacuum fluctuations.
A dashed line indicates a second-order transition, while a 
solid line indicates a first-order transition. 
The region between the red lines is the inhomogeneous phase.}
\label{chiral2}
\end{figure}

In Fig. \ref{chiral2}, we show the phase diagram in the $\mu$--$T$ plane 
in the chiral limit where vacuum fluctuations are included.
The inhomogeneous phase in the entire $\mu$--$T$ plane 
has now been replaced by a small region at low temperatures.
The second-order line starting at $\mu=0$ ends at the Lifshitz
point indicated by the full red circle. 
Since $m_\sigma=2m_q$
this is also the position of the tricrital point \cite{nick2}.
The region between the
two red lines is the inhomogeneous phase.
Comparing Figs. \ref{chiral} and \ref{chiral2}, we see the dramatic
effects of including the fermionic vacuum fluctuations.

$ $\\

Although we find an inhomogeneous phase for finite temperature, it has to be mentioned that this phase might not survive if effects beyond the mean-field approximation are included. There is evidence that in the chiral limit the existence of the Lifshitz point is simply an artifact if the mean-field approximation as pointed out in Ref. \cite{japs} and Ref. \cite{diehl}.

In Ref. \cite{japs}, it is shown that the Goldstone bosons
that arise from the breaking of the translational and rotational symmetry
have a quadratic dispersion relation in some directions and a linear
dispersion relation in other directions. 
At finite temperature, the former leads
to strong long-wavelength fluctuations (phase fluctuations) that 
destroy off-diagonal long-range order altogether.
Long-range order is replaced by quasi-long-range order 
where the order parameter is decaying algebraically.
At $T=0$, the phase fluctuations are not strong enough to 
destroy this order and there is a true condensate.

In Fig. \ref{gap}, we show the modulus $\Delta$ (solid blue line)
and the wave vector
$q$ (dashed red line) as functions of $\mu$ at $T=0$ in the chiral limit with
$m_{\sigma}=2m_q=600$ MeV. The left panel shows the results
without quantum fluctuations and the right panel with.
The transition from a phase
with homogeneous condensate to a phase with a 
chiral-density wave is first order, while the transition to a
chirally symmetric phase is second order.
In the case with no vacuum fluctuations, the vacuum 
state, i.e. with zero quark density extends all the way to
the transition to the CDW phase which extends from $\mu=291\;$MeV up to $\mu=384\;$MeV. This is not the case if we include
quantum fluctuations. The vacuum state extends from $\mu=0$
up to $\mu=291$ MeV, where
there is a transition to a homogeneous phase with a nonzero
quark density and $\Delta$ decreases. This phase 
extends up to $\mu\approx322.7$ MeV.
In both cases, the vanishing quark density for $\mu<\mu_c$, where
$\mu_c$ is the critical density for the transition to either the CDW phase (left panel) or another homogeneous phase (right panel) with decreasing $\Delta$
is an example of the silver-blaze property. In this phase, all physical
quantities are independent of the quark chemical potential \cite{cohen}.

\begin{widetext}

\begin{figure}[htb]
\begin{center}
\includegraphics[width=0.4\textwidth]{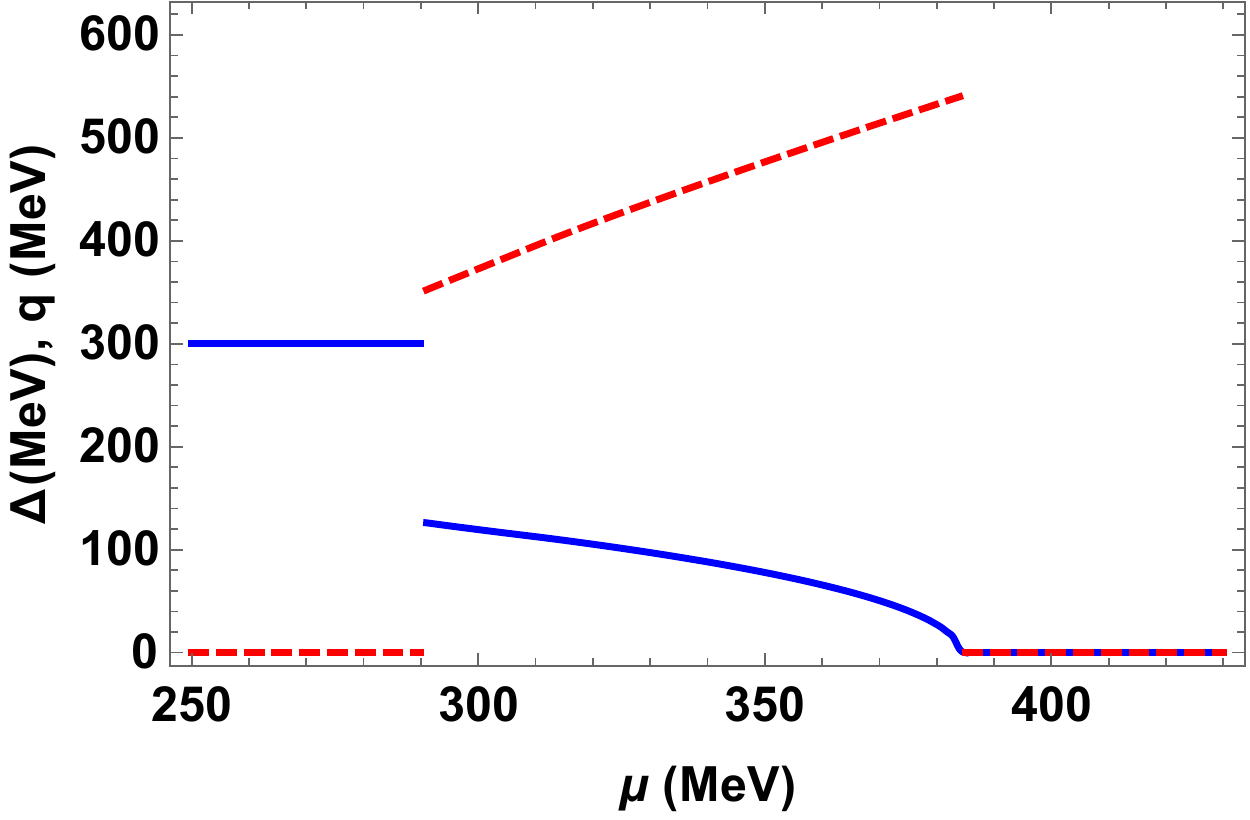}
\includegraphics[width=0.4\textwidth]{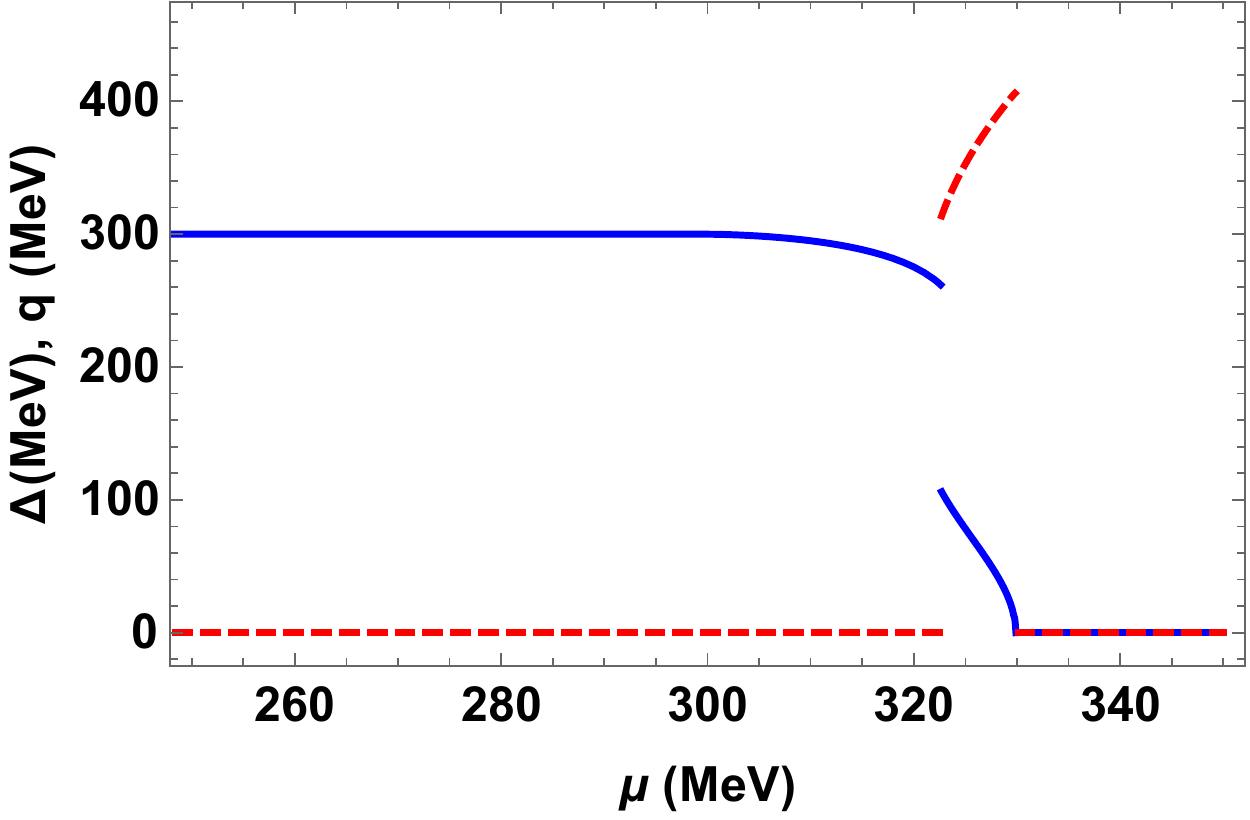}
\end{center}
\caption{ Gap $\Delta$ (solid blue line)
and wavevector $q$ (dashed red line)
as functions of the quark chemical potential $\mu$
in the chiral limit, 
at $T=0$, and for $m_{\sigma}=2m_q=600$ MeV.
Left panel is without vacuum fluctuations and right panel 
with vacuum fluctuations.}
\label{gap}
\end{figure}

In Fig. \ref{gap2}, we 
show the modulus $\Delta$ (solid blue line)
and the wave vector
$q$ (dashed red line) as functions of $\mu$ at $T=0$ at the physical point with
$m_{\sigma}=2m_q=600$ MeV and $m_\pi=140$ MeV.
In the left panel, we have omitted the quantum fluctuations and in the
right panel, they  have been included.
Without quantum corrections, there is a transition from
a phase with a homogeneous chiral condensate to a phase with 
a chiral-density wave. This transition is first order.
Again, this is in contrast to the case where we include the vacuum fluctuations;
the vacuum phase extends from $\mu=0$ to $\mu=300$ MeV and then a second order transition occurs to a phase with a homogeneous quark chiral condensate and a nonzero
quark density. In this phase, the chiral condensate decreases.
There are two more transitions, 
one from the phase with a homogeneous
chiral condensate (and a nonzero quark density)
to a phase with an inhomogeneous phase and a
transition to a chirally symmetric phase.
Both transitions are first order.


\begin{figure}[htb]
\begin{center}
\includegraphics[width=0.4\textwidth]{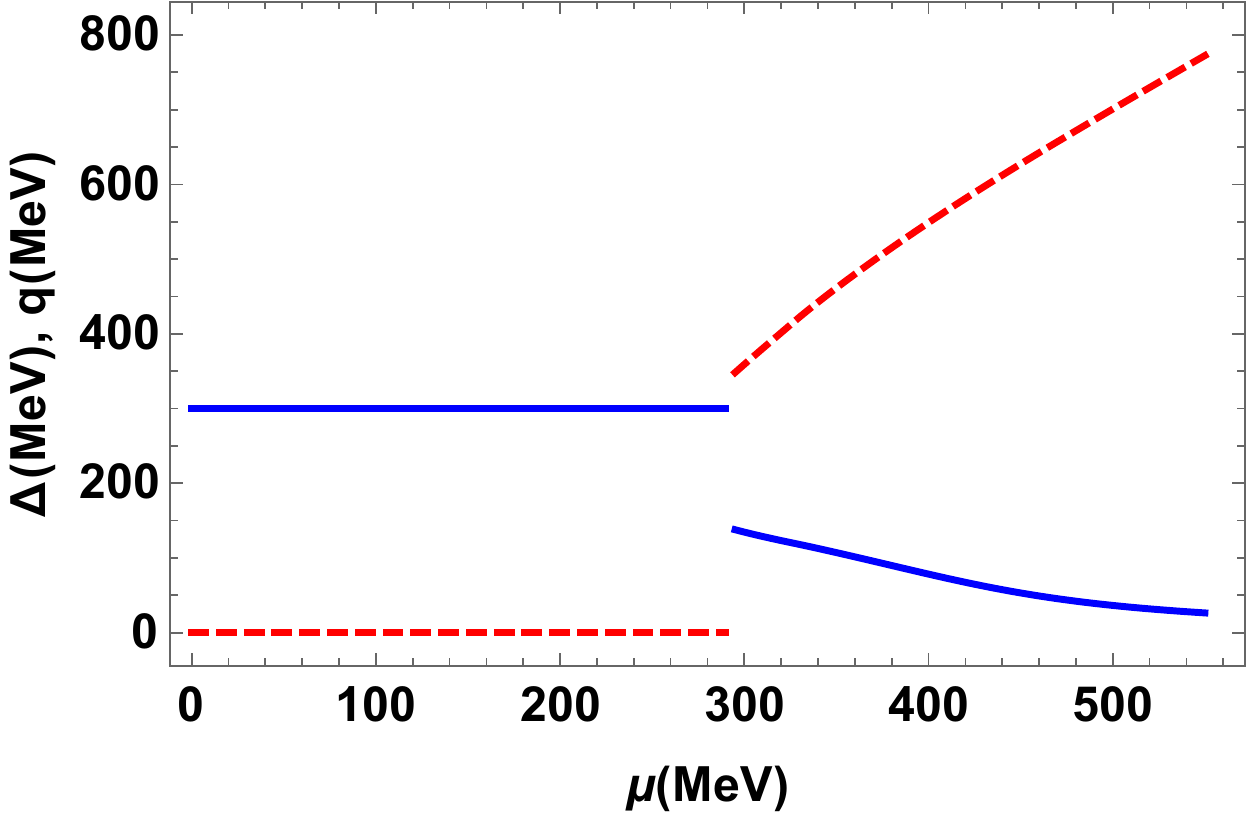}
\includegraphics[width=0.4\textwidth]{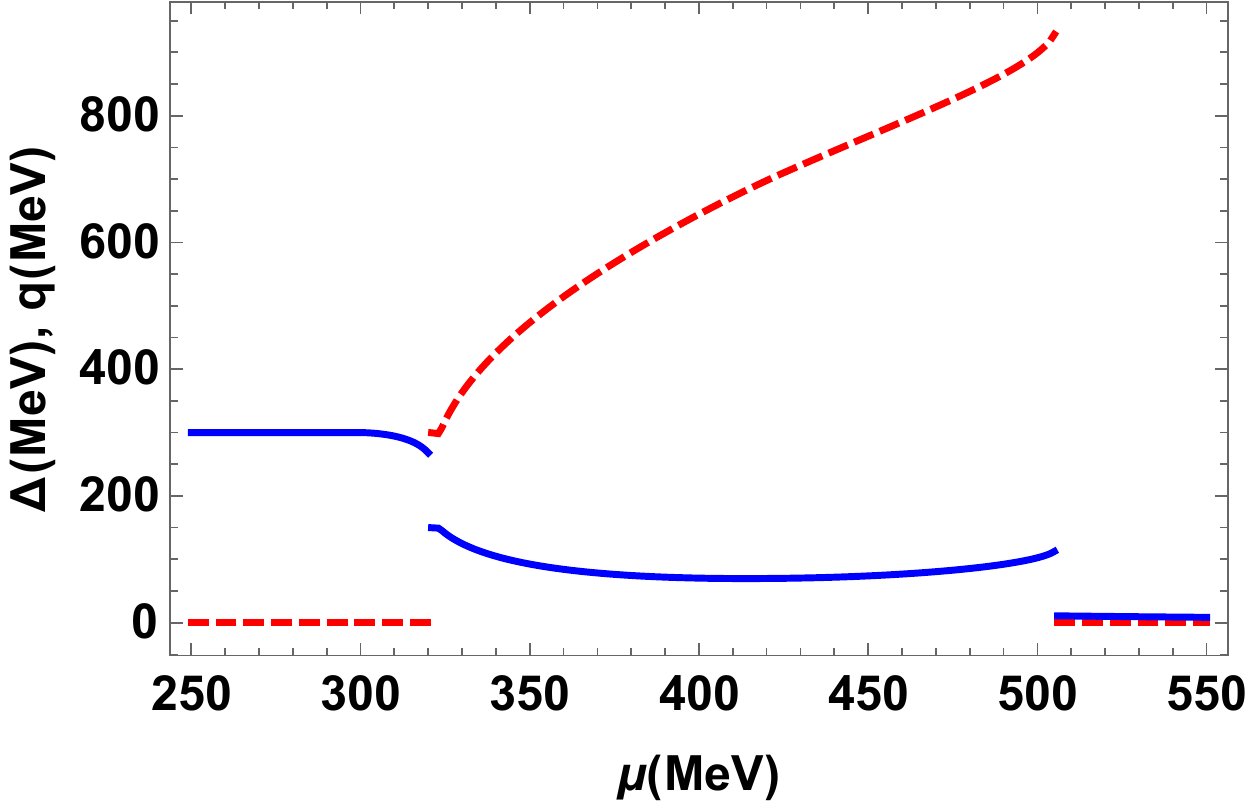}
\end{center}
\caption{Gap $\Delta$ (solid blue line)
and wave vector $q$ (dashed red line)
as functions of the quark chemical potential $\mu$
at the physical point for $T=0$ and $m_{\sigma}=2m_q=600$ MeV.
Left panel is without vacuum fluctuations and right panel 
with vacuum fluctuations.
}
\label{gap2}
\end{figure}

\end{widetext}

The present work can be extended in different directions.
For example, it would be of interest to study inhomogeneous phases
in a constant magnetic background. Work in this direction is in 
progress \cite{ankn}.


\section*{ACKNOWLEDGMENTS}
The authors would like to thank Stefano Carignano for useful discussions.
P.A. would like to acknowledge the research travel support provided through 
the Professional Development Grant and would like to thank the Faculty Life 
Committee and the Dean's Office at St. Olaf College. P.A. would also like to 
acknowledge the computational support provided through the Computer Science 
Department at St. Olaf College and thank Richard Brown, Tony Skalski and 
Jacob Caswell. 
P.A. and P.K. would like to thank the Department of
Physics at NTNU for kind hospitality during the latter stages of this work.

\appendix

\section{INTEGRALS AND SUM INTEGRALS}
\label{integrali}
In the imaginary-time formalism for thermal field theory, a 
fermion has Euclidean 4-momentum $P=(P_0,{\bf p})$ with
$P^2=P_0^2+{\bf p}^2$. The Euclidean energy $P_0$
has discrete values:
$P_0=(2n+1)\pi T+i\mu$, where $n$ is an
integer. Loop diagrams involve a sum over
$P_0$ and an integral over spatial momenta $p$. 
We define the dimensionally regularized
sum integral by
\bqa
\sumint_{\{P\}}&=&
T\sum_{P_0}\int_p\;,
\label{sumint}
\eqa
where the integral is in $d=3-2\epsilon$ dimensions
\bqa\nonumber
\int_p&=&\left({e^{\gamma_E}\Lambda^2\over4\pi}\right)^{\epsilon}
\int{d^dp\over(2\pi)^d}
\\ &=& \nonumber
\left({e^{\gamma_E}\Lambda^2\over4\pi}\right)^{\epsilon}
\int{d^{d-1}p_{\perp}\over(2\pi)^{d-1}}
\int{dp_{\parallel}\over2\pi}
\\&=&
\left({e^{\gamma_E}\Lambda^2\over4\pi}\right)^{\epsilon}
\int{d^{d-1}p_{\perp}\over(2\pi)^{d-1}}
{1\over \pi}\int_\Delta^\infty{u\,du\over\sqrt{u^2-\Delta^2}}
\label{defint}
\;.
\eqa
Here $\Lambda$ is the renormalization scale in the
modified minimal subtraction scheme $\overline{\rm MS}$, 
$p_{\parallel}=p_3$, $p_{\perp}^2=p_1^2+p_2^2$ and $u=\sqrt{p_\parallel^2+\Delta^2}$.
We need
\bqa
I_0&=&-\sumint_{\{P\}}\log\left[
P_0^2+E^2
\right]\;.
\eqa
Summing over the Matsubara frequencies $P_0$, we obtain
\bqa\nonumber
I_0&=&
-\int_p\bigg\{E
+T\log\left[1+e^{-\beta(E-\mu)}\right]
\\ &&
+T\log\left[1+e^{-\beta(E+\mu)}\right]\bigg\}\;.
\label{simint}
\eqa
The integral in Eq. (\ref{simint}) is needed for $E=E_{\pm}$ and
is calculated by expanding it in powers of $q$, see Appendix \ref{epder}.
The integrals that appear are
\bqa\nonumber
\int_p\sqrt{u^2+p^2_{\perp}}
&=&
-{\Delta^4\over(4\pi)^2}\left({e^{\gamma_E}\Lambda^2\over\Delta^2}\right)^{\epsilon}
\Gamma(-2+\epsilon)\;,
\\ &=&
-{\Delta^4\over 2(4\pi)^2}\left({\Lambda^2\over\Delta^2}\right)^{\epsilon}
\left[{1\over\epsilon}+{3\over2}+{\cal O}(\epsilon)
\right]
\;,
\\ \nonumber
\int_p{p_{\perp}^2\over(u^2+p^2_{\perp})^{3\over2}}
&=&
-{4\Delta^2\over(4\pi)^2}\left({e^{\gamma_E}\Lambda^2\over\Delta^2}\right)^{\epsilon}
\Gamma(\epsilon)
\\
&=&
-{4\Delta^2\over(4\pi)^2}\left({\Lambda^2\over\Delta^2}\right)^{\epsilon}
\left[{1\over\epsilon}
+{\cal O}(\epsilon)
\right]\;,
\\ \nonumber
\int_p{p_{\perp}^2(4u^2-p_{\perp}^2)\over(u^2+p^2_{\perp})^{7\over2}}
&=&
-\dfrac{16}{3(4\pi)^2}\left({e^{\gamma_E}\Lambda^2\over\Delta^2}\right)^{\epsilon}
(-1+\epsilon)\Gamma(1+\epsilon)
\\ &=&
{1\over3\pi^2} +\mathcal{O}(\epsilon)\;.
\eqa
We also need some integrals in $D=4-2\epsilon$ dimensions
Specifically, we need the integrals
\bqa\nonumber
A(m^2)&=&
\int_p{1\over p^2-m^2}
\label{defa}
\\&=& 
{im^2\over(4\pi)^2}\left({\Lambda^2\over m^2}\right)^{\epsilon}
\left[{1\over\epsilon}+1+\mathcal{O}(\epsilon)\right]\;,
\\ \nonumber
B(p^2)&=&
\int_k{1\over(k^2-m_q^2)[(k+p)^2-m_q^2]}
\\ &=& 
{i\over(4\pi)^2}\left({\Lambda^2\over m_q^2}\right)^{\epsilon}\left[
{1\over\epsilon}+F(p^2)+\mathcal{O}(\epsilon)
\right]\;,
\label{defb}
\\
B^{\prime}(p^2)&=&{i\over(4\pi)^2}F^{\prime}(p^2)\;,
\eqa
where the functions $F(p^2)$ and $F^{\prime}(p^2)$ are
\bqa
\\ \nonumber
F(p^2)&=& 
-\int_0^1dx\,\log\left[{p^2\over m_q^2}x(x-1)+1\right] 
\\&=&
2-2r\,{\arctan}\left(\mbox{$1\over r$}\right)
\;,
\label{fdep}
\\
F^{\prime}(p^2)&=&
{4m_q^2r\over p^2(4m_q^2-p^2)}{\arctan}
\left(\mbox{$1\over r$}\right)
-{1\over p^2}
\;,
\label{defbb}
\eqa
were we defined $ r= \sqrt{{4m_q^2\over p^2}-1}$.

\section{PARAMETER FIXING}
\label{appfix}
In this appendix, we find the relation between the parameters in the
Lagrangian (\ref{lag}) and the physical observables using the
$\overline{\rm MS}$ and OS renormalization schemes.

The sigma and pion self-energies are given by
\bqa\nonumber
\Sigma_{\sigma}(p^2)&=&
-8g^2N_c
\left[A(m_q^2)-\mbox{$1\over2$}(p^2-4m_q^2)
B(p^2)
\right]
\\ &&
+{4\lambda g\phi_0N_cm_q\over m_{\sigma}^2}A(m_q^2)
\;,
\label{sigsum}
\\ \nonumber
\Sigma_{\pi}(p^2)&=&
-8g^2N_c\left[A(m_q^2)-\mbox{$1\over2$}p^2B(p^2)\right]
\\ 
&&
+{4\lambda g\phi_0N_cm_q\over3m_{\sigma}^2}A(m_q^2)\;,
\label{pisum}
\eqa
where the last term of Eqs. (\ref{sigsum}) and
(\ref{pisum}) is the tadpole contribution to the self-energies, and 
where the integrals $A(m^2)$ and $B(p^2)$ are defined in 
Eqs. (\ref{defa}) and (\ref{defb}).
We do not need the quark self-energy since it is of order $N_c^0$.
Thus $Z_{\psi}=1$ and $\delta m_q=0$ at this order.
The inverse propagator for the sigma or pion 
can be written as
\bqa
p^2-m_{\sigma,\pi}^2-i\Sigma_{\sigma,\pi}(p^2) +
{\rm counterterms}
\;.
\label{definv}
\eqa
In the on-shell scheme, the physical mass 
is equal to the renormalized mass in the 
Lagrangian.\footnote{In defining the mass, we ignore the imaginary parts
of the self-energy.}
Thus we can write
\bqa
\Sigma_{\sigma,\pi}(p^2=m_{\sigma,\pi}^2)
{\rm + counterterms}
&=&0
\label{pole}
\;.
\eqa
The residue of the propagator on shell equals unity, which implies
\bqa
\label{res}
{\partial\over\partial p^2}\Sigma_{\sigma,\pi}(p^2)\big|_{p^2=m_{\sigma,\pi}^2}
{\rm +counterterms}
&=&0\;.
\eqa
The large-$N_c$ contribution to the one-point function is
\bqa
\delta\Gamma^{(1)}
&=&-8g^2N_c \phi_0 A(m_q^2)+i\delta t
\label{dg1}
\;,
\eqa
where $\delta t$ is the tadpole counterterm.
The equation of motion is equivalent to the vanishing one-point function, which yields on tree level $t=h-m_{\pi}^2\phi_0=0$.
This has to hold also on one-loop level, which gives the renormalization condition
\bqa
\delta\Gamma^{(1)}
&=&0\;.
\label{dg2}
\eqa
The counterterms are given by
\bqa
\label{count1}
\Sigma_{\sigma}^{\rm ct1}(p^2)
&=&i\left[
\delta Z_{\sigma}(p^2-m_{\sigma}^2)-\delta m_{\sigma}^2
\right]\;,\\
\label{count2}
\Sigma_{\pi}^{\rm ct1}(p^2)
&=&i\left[
\delta Z_{\pi}(p^2-m_{\pi}^2)-\delta m_{\pi}^2
\right]\;,
\\ 
\label{cont2}
\Sigma_{\sigma}^{\rm ct2}&=&3\Sigma_{\pi}^{\rm ct2}
=-{i\lambda\phi_0\over m_{\sigma}^2}\delta t\;,
\\\delta t&=&\delta h-
f_{\pi}\delta m_{\pi}^2-m_{\pi}^2\delta f_{\pi} 
\;,
\label{dt11}
\eqa
where the counterterm in Eq. (\ref{cont2}) cancels the
tadpole contribution to the self-energies.
The on-shell renormali\-zation constants are given by the
self-energies and their derivatives 
evaluated at the physical mass.
This yields
\bqa
\delta m_{\sigma}^2&=&-i\Sigma_{\sigma}(m_{\sigma}^2)\;,\\
\delta m_{\pi}^2&=&-i\Sigma_{\pi}(m_{\pi}^2)\;,\\
\delta Z_\sigma &=&
i{\partial\over\partial p^2}\Sigma_\sigma(p^2)|_{p^2=m_\sigma^2}\;, \\
\delta Z_\pi&=& i{\partial\over\partial p^2}\Sigma_\pi(p^2)|_{p^2=m_\pi^2}
\;.
\eqa
From Eqs. (\ref{sigsum})--(\ref{dg1}), 
we find \footnote{The self-energies are without the tadpole contributions.}
\bqa
\delta m_{\sigma}^2
&=&
8ig^2N_c\left[A(m_q^2)-\mbox{$1\over2$}(m_{\sigma}^2-4m_q^2)B(m_{\sigma}^2)
\right]\;,
\\ 
\delta m_{\pi}^2&=&
8ig^2N_c\left[A(m_q^2)-\mbox{$1\over2$}m_{\pi}^2B(m_{\pi}^2)\right]\;,
\\ 
\delta Z_{\sigma}&=&
4ig^2N_c\left[B(m_{\sigma}^2)+(m_{\sigma}^2-4m_q^2)B^{\prime}(m_{\sigma}^2)
\right]\;,
\\
\delta Z_{\pi}&=&4ig^2N_c
\left[
B(m_{\pi}^2)+m_{\pi}^2B^{\prime}(m_{\pi}^2)
\right]\;,\\
\delta t&=&-8ig^2N_c f_\pi A(m_q^2)
\;.
\eqa
The counterterms $\delta m^2$, $\delta\lambda$, $\delta g^2$, and 
$\delta h$ can be expressed in terms of the counterterms
$\delta m_{\sigma}^2$, $\delta m_{\pi}^2$, $\delta Z_{\pi}$, and $\delta t$.
Since there is no correction to the quark-pion vertex
in the large-$N_c$ limit, we find 
\bqa
\delta g^2=-g^2\delta Z_{\pi}\;.
\eqa
Since there is no correction to the quark mass 
in the large-$N_c$ limit, we find $\delta m_q=0$ or
\bqa
\delta g^2&=&-g^2{\delta f_{\pi}^2\over f_{\pi}^2}\;.
\eqa
This yields $\delta Z_{\pi}={\delta f_{\pi}^2\over f_{\pi}^2}$.
From this relation, Eq. (\ref{dt11}), and $h=t+m_{\pi}^2f_{\pi}$, 
one finds
\bqa
\delta m^2&=&-{1\over2}\left(\delta m_{\sigma}^2-3\delta m_{\pi}^2\right)\;,
\label{rela1}
\\
\delta\lambda&=&
3{\delta m_{\sigma}^2-\delta m_{\pi}^2\over f_{\pi}^2}
-\lambda\delta Z_{\pi}
\;,
\label{rela3}
\\
\delta h&=&\delta t+ f_{\pi}\delta m_{\pi}^2+{1\over2}m_{\pi}^2f_{\pi}\delta Z_{\pi}
\;.
\eqa
\begin{widetext}
The expressions for the counterterms are
\bqa\nonumber
\delta m_{\os}^2&=&
8ig^2N_c\left[A(m_q^2)+\mbox{$1\over4$}(m_{\sigma}^2-4m_q^2)B(m_{\sigma}^2)
-\mbox{$3\over4$}m_{\pi}^2B(m_{\pi}^2)
\right]
\\&=&\delta m^2_{\rm div} +
\dfrac{4g^2 N_c}{(4\pi)^2} 
\left\{ m^2\log\mbox{$\Lambda^2\over m_q^2$}
-2m_q^2
-{1\over2}\left(m_\sigma^2-4m_q^2\right) F(m_\sigma^2) +{3\over2}m_{\pi}^2F(m_{\pi}^2)
\right\}  
\;,
\label{osse1}
\\ \nonumber
\delta\lambda_{\os}&=&
-\dfrac{12ig^2N_c}{f_{\pi}^2} (m_\sigma^2-4m_q^2)B(m_\sigma^2) 
+\dfrac{12ig^2N_c}{f_{\pi}^2}m_\pi^2B(m_\pi^2) 
-4i\lambda g^2N_c\left[B(m_{\pi}^2)+m_{\pi}^2B^{\prime}(m_{\pi}^2)\right]
\\ &=& \nonumber
\delta\lambda_{\rm div} + \dfrac{12g^2 N_cm_\sigma^2}{(4\pi)^2f_\pi^2}
\left[\left(1-{4m_q^2\over m_\sigma^2}\right)
\left[\log\mbox{$\Lambda^2\over m_q^2$}
+F(m_\sigma^2)
\right] 
+\log\mbox{$\Lambda^2\over m_q^2$}
+F(m_\pi^2)+m_\pi^2F^{\prime}(m_\pi^2) \right]
\\
&&
-\dfrac{12g^2 N_cm_\pi^2}{(4\pi)^2f_\pi^2}
\left[
2\log\mbox{$\Lambda^2\over m_q^2$}
+2F(m_\pi^2)
+m_\pi^2F^{\prime}(m_\pi^2)\right]
\;,
\label{osse2}
\\ 
\delta g^2_{\os}&=&
-4ig^4N_c\left[B(m_{\pi}^2)+m_{\pi}^2B^{\prime}(m_{\pi}^2)\right]
=\delta g_{\rm div}^2
+\dfrac{4g^4 N_c}{(4\pi)^2}\left[ 
\log\mbox{$\Lambda^2\over m_q^2$}
+F(m_\pi^2) +m_\pi^2F^\prime(m_\pi^2) \right]
\;,
\label{osse3}
\\ \nonumber
\delta h^{\os}
&=&-2ig^2N_cm_\pi^2f_\pi\left[
B(m_\pi^2)-m_\pi^2B^{\prime}(m_\pi^2)\right]
\\&=&
\delta h_{\rm div}+{2g^2N_cm_\pi^2f_\pi\over(4\pi)^2}
\left[
\log\mbox{$\Lambda^2\over m_q^2$}+F(m_\pi^2)-m_\pi^2F^{\prime}(m_\pi^2)
\right]
\;,
\label{osse4}
\\
\delta Z_{\sigma}^{\os}
&=&
\delta Z_{\sigma,\rm div}
-\dfrac{4g^2 N_c}{(4\pi)^2}\left[ 
\log\mbox{$\Lambda^2\over m_q^2$}
+F(m_\sigma^2) +(m_\sigma^2-4m_q^2)F^\prime(m_\sigma^2) \right]
\;\;\;, \\
\delta Z_{\pi}^{\os}
&=&
\delta Z_{\pi,\rm div}
-\dfrac{4g^2N_c}{(4\pi)^2}\left[
\log\mbox{$\Lambda^2\over m_q^2$}
+F(m_\pi^2) +m_\pi^2F^\prime(m_\pi^2) \right]\;,
\eqa
where $F(m^2)$ and $F^{\prime}(m^2)$
are defined in Appendix \ref{integrali}, and 
the divergent quantities are
\bqa
\delta m^2_{\rm div}&=&
{4m^2g^2N_c \over(4\pi)^2\epsilon}
\;,
\hspace{1cm}
\delta\lambda_{\rm div}=
{8N_c\over(4\pi)^2\epsilon}
\left(\lambda g^2-6g^4
\right)
\;,\hspace{1cm}
\delta g_{\rm div}^2=
{4g^4N_c\over(4\pi)^2\epsilon}\;,\\
\delta Z_{\sigma,\rm div}&=&\delta Z_{\pi,\rm div}=-{4g^2N_c\over(4\pi)^2\epsilon}
\;,\hspace{1cm}
\delta h_{\rm div}
{2g^2hN_c\over(4\pi)^2\epsilon}
\;.
\eqa
\end{widetext}
The divergent parts of the counterterms 
are the same in the two schemes, i.e.
$\delta m^2_{\rm div}=\delta m^2_{\ms}$ and so forth.
Since the bare parameters are independent
of the renormalization scheme, we can immediately
write down relations between
the renormalized parameters in the on-shell and $\overline{\rm MS}$ schemes.
We find
\begin{figure}
\end{figure}
\bqa
m^2_{\ms} &=& 
m^2 + \delta m^2_{\os} 
- \delta m^2_{\ms} 
\\
\lambda_{\ms} &=& 
\lambda + \delta\lambda_{\os}  - \delta\lambda_{\ms} 
\\
g_{\ms}^2 &=& 
g^2 + \delta g^2_{\os}- \delta g^2_{\ms}\;,
\\
h_{\ms}&=&
h+\delta h_{\os}-\delta h_{\ms}\;.
\eqa
Using Eqs. (\ref{osse1})--(\ref{osse4}), we find the 
running parameters in the $\overline{\rm MS}$ scheme
\begin{widetext}
\bqa \nonumber
m^2_{\ms} 
&=& m^2 +8ig^2N_c \left[ A(m_q^2) +\mbox{$1\over4$}(m_\sigma^2-4m_q^2)
B(m_\sigma^2) 
-\mbox{$3\over4$}m_{\pi}^2B(m_{\pi}^2)
\right] -\delta m^2_{\ms} \\
 &=& m^2+
\dfrac{4g^2N_c}{(4\pi)^2} 
\left[m^2\log\mbox{$\Lambda^2\over m_q^2$}
-2m_q^2-{1\over2}\left(m_\sigma^2-4m_q^2\right)F(m_\sigma^2) 
+{3\over2}{m_\pi^2}F(m_\pi^2) 
\right]\;, 
\label{osm1}
\\ \nonumber
\lambda_{\ms} 
&=& \lambda 
-\dfrac{12ig^2N_c}{f_{\pi}^2} (m_\sigma^2-4m_q^2)B(m_\sigma^2) 
+\dfrac{12ig^2N_c}{f_{\pi}^2}m_\pi^2B(m_\pi^2) 
-4i\lambda g^2N_c\left[B(m_{\pi}^2)+m_{\pi}^2B^{\prime}(m_{\pi}^2)\right]
- \delta\lambda_{\ms}
\\ \nonumber &=& 
\lambda
+\bigg\{\dfrac{12g^2N_c}{(4\pi)^2f_\pi^2}
\left[(m_\sigma^2-4m_q^2)\left(
\log\mbox{$\Lambda^2\over m_q^2$} +F(m_\sigma^2)
\right)
+m_\sigma^2\left(
\log\mbox{$\Lambda^2\over m_q^2$}
+F(m_\pi^2)
+m_\pi^2F^{\prime}(m_\pi^2)
\right)
\right.
\\ &&
\left.
-m_\pi^2\left(
2\log\mbox{$\Lambda^2\over m_q^2$}+2F(m_\pi^2)
+F^{\prime}(m_\pi^2)\right)\right]
\bigg\}\;,
\label{osl}
\\ \nonumber
g_{\ms}^2  &= &g^2-4ig^4N_c
\left[B(m_\pi^2) +m_\pi^2B^\prime(m_\pi^2) \right] -\delta g_{\ms}^2 
\\ 
 &=& g^2
\left\{1 + \dfrac{4g^2N_c}{(4\pi)^2}\left[
\log\mbox{$\Lambda^2\over m_q^2$}
+F(m_\pi^2) +m_\pi^2F^\prime(m_\pi^2)
\right]\right\}
\;,
\label{g00}
\\ \nonumber
h_{\ms}&=&h-
2ig^2N_c h\left[
B(m_\pi^2)-m_\pi^2B^{\prime}(m_\pi^2)
\right]-\delta h_{\ms}
\\
&=&h+{2g^2 N_c h\over(4\pi)^2}
\left[\log\mbox{$\Lambda^2\over m_q^2$}
+F(m_\pi^2)-m_\pi^2F^{\prime}(m_\pi^2)
\right]
\;,
\label{hhh}
\eqa
\end{widetext}
where the physical on-shell values are related to the
meson and quark masses given by Eqs. (\ref{tr1})--(\ref{tr4}).

\section{EFFECTIVE POTENTIAL}
\label{epder}
In this appendix, we calculate the one-loop effective potential
in the $\overline{\rm MS}$ scheme. 
It reads
\bqa\nonumber
V_1&=&-2N_c\sumint_{\{P\}}\log[P_0^2+E_{\pm}^2]
\\ \nonumber
&=&-2N_c\int_p\bigg\{E_{\pm}
+T\log\left[1+e^{-\beta(E_{\pm}-\mu)}\right]
\\ &&
+T\log\left[1+e^{-\beta(E_{\pm}+\mu)}\right]\bigg\}
\;,
\eqa 
where the sum integral is defined in Eq. (\ref{sumint}).
The vacuum integrals needed are
\bqa
V_{\pm}&=&-2N_c\int_pE_{\pm}\;.
\label{intv}
\eqa
We first integrate over angles in the $(p_1,p_2)$ plane 
and introduce the variable
$u=\sqrt{p_{\parallel}^2+\Delta^2}$.
The expression for $V_{\pm}$ can then be written as
\bqa\nonumber
V_{\pm}&=&-{16 N_c(e^{\gamma_E}\Lambda^2)^{\epsilon}\over(4\pi)^{2}\Gamma(1-\epsilon)}
\int_{\Delta}^{\infty}
{u\,du\over\sqrt{u^2-\Delta^2}}
\\ &&
\times
\int_{0}^{\infty}
\sqrt{\left(u\pm{q\over2}\right)^2+p_{\perp}^2}\,p_{\perp}^{1-2\epsilon}dp_{\perp}
\;.
\label{integralane}
\eqa
The strategy is to isolate the ultraviolet divergences 
in Eq. (\ref{integralane}) by expanding the integrand 
and identifying appropriate subtraction terms ${\rm sub}_{\pm}(u,p_{\perp})$.
The integral of the subtraction terms 
can be done in dimensional regularization, 
while the integral of $E_{\pm}-{\rm sub}_{\pm}(u,p_{\perp})$
is finite and can be calculated directly in three dimensions.
The subtraction term ${\rm sub}_{\pm}(u,p_{\perp})$
is found by expanding Eq. (\ref{integralane}) through order $q^4$.
This yields
\begin{widetext}
\bqa
{\rm sub}_{\pm}(u,p_{\perp})
&=&
\sqrt{u^2+p^2_{\perp}}\pm{uq\over2\sqrt{u^2+p_{\perp}^2}}
+{q^2p_{\perp}^2\over8(u^2+p_{\perp}^2)^{3\over2}}
\mp{q^3p_{\perp}^2u\over16(u^2+p_{\perp}^2)^{5\over2}}
+{q^4p_{\perp}^2(4u^2-p_{\perp}^2)\over128(u^2+p_{\perp}^2)^{7\over2}}\;.
\label{sub22}
\eqa
We write the integrals in (\ref{integralane}) as
\bqa
V_{\pm}&=&V_{\rm div\pm}+V_{\rm fin\pm}\;,
\eqa
where 
\bqa
V_{\rm div\pm}&=&
-{16N_c(e^{\gamma_E}\Lambda^2)^{\epsilon}\over(4\pi)^{2}\Gamma(1-\epsilon)}
\int_{\Delta}^{\infty}
{u\,du\over\sqrt{u^2-\Delta^2}}
\int_{0}^{\infty}
{\rm sub}_{\pm}(u,p_{\perp})p_{\perp}^{1-2\epsilon}\,dp_{\perp}
\;,
\\
V_{\rm fin\pm}&=&
-{16N_c(e^{\gamma_E}\Lambda^2)^{\epsilon}\over(4\pi)^{2}\Gamma(1-\epsilon)}
\int_{\Delta}^{\infty}
{u\,du\over\sqrt{u^2-\Delta^2}}
\int_{0}^{\infty}
\left[\sqrt{\left(u\pm{q\over2}\right)^2+p_{\perp}^2}
-{\rm sub}_{\pm}(u,p_{\perp})\right]
p_{\perp}^{1-2\epsilon}\,dp_{\perp}\;.
\eqa
The integral $V_{\rm fin\pm}$ can now be calculated directly in three dimensions.
After integrating over $p_{\perp}$, we find
\bqa
V_{\rm fin\pm}&=&
-{16N_c
\over3(4\pi)^{2}}
\int_{\Delta}^{\infty}
(u\pm\mbox{$q\over2$})^2\left[
\left(u\pm\mbox{$q\over2$}\right)-\big|u\pm\mbox{$q\over{2}$}\big|
\right]{u\,du\over\sqrt{u^2-\Delta^2}}\;.
\eqa
Thus $V_{\rm fin+}$ vanishes identically
and $V_{\rm fin-}$ becomes
\bqa \nonumber
V_{\rm fin -}
&=&
-{32N_c\over3(4\pi)2^2}\int_{\Delta}^{\infty}
(u-\mbox{$q\over2$})^3
\theta(\mbox{$q\over2$}-\Delta){u\,du\over\sqrt{u^2-\Delta^2}}
\\ 
&=&
{N_c\over3(4\pi)^2}
\left[q\sqrt{{q^2\over4}-\Delta^2}(26\Delta^2+q^2)
-12\Delta^2(\Delta^2+q^2)\log{{q\over2}+\sqrt{{q^2\over4}-\Delta^2}\over\Delta}
\right]
\theta(\mbox{$q\over2$}-\Delta)
\;.
\label{finf}
\eqa
We next  integrate $V_{\rm div\pm}$ using dimensional regularization.
Again this is done by first integrating over $p_{\perp}$ and then over $u$.
This yields
\bqa\nonumber
V_{\rm div}&=&V_{\rm div+}+V_{\rm div-}
\\
&=&
{2N_c\over(4\pi)^2}
\left({e^{\gamma_E}\Lambda^2\over\Delta^2}\right)^{\epsilon}
\left[
2\Delta^4
\Gamma(-2+\epsilon)+q^2\Delta^2\Gamma(\epsilon)
+{q^4\over12}(-1+\epsilon)\Gamma(1+\epsilon)
\right]\;.
\label{vd}
\eqa
Expanding Eq. (\ref{vd}) to zeroth order in powers of $\epsilon$, we obtain
\bqa
V_{\rm div}&=&
{2N_c\over(4\pi)^2}\left({\Lambda^2\over\Delta^2}
\right)^{\epsilon}
\left[
\left({1\over\epsilon}+{3\over2}\right)\Delta^4
+{1\over\epsilon}\Delta^2q^2
-{q^4\over12}
+{\cal O}(\epsilon)
\right]\;.
\label{divf}
\eqa
The one-loop effective potential is then given by the sum of Eqs. (\ref{finf})
and (\ref{divf}). It 
contains poles in $\epsilon$, which are
removed by mass and coupling-constant renormalization.
In the $\overline{\rm MS}$ scheme,
this  amounts to making the substitutions
$m^2\rightarrow Z_{m^2}m^2$, 
$\lambda\rightarrow Z_{\lambda}\lambda$, $g^2\rightarrow Z_{g^2}g^2$, and
$h\rightarrow Z_{h}h$, 
where 
\bqa
Z_{m^2}=1+{4N_cg^2\over(4\pi)^2\epsilon}\;,
\hspace{0.8cm}
Z_{\lambda}=1+
{8N_c\over(4\pi)^2\epsilon}
\left[
\lambda g^2-6g^4\right]
\;,
\label{dl}
\hspace{0.8cm}Z_{g^2}=1+{4N_cg^2\over(4\pi)^2\epsilon}
\hspace{0.8cm}Z_{h}=1+{2N_cg^2\over(4\pi)^2\epsilon}
\;.
\eqa
After renormalization, the vacuum
energy in the mean-field approximation reads
\bqa\nonumber
V_++V_-
&=&
{1\over2}{q^2\over g_{\ms}^2(\Lambda^2)}\Delta^2
+{1\over2}{m_{\ms}^2(\Lambda^2)\over g_{\ms}^2(\Lambda^2)}\Delta^2
+{\lambda_{\ms}(\Lambda^2)\over24g_{\ms}^4(\Lambda^2)}\Delta^4
-{h_{\ms}(\Lambda^2)\over g_{\ms}(\Lambda^2)}\Delta
+{2N_c\Delta^2q^2\over(4\pi)^2}
\log{\Lambda^2\over\Delta^2}
+{2N_c\Delta^4\over(4\pi)^2}\left[
\log{\Lambda^2\over\Delta^2}+{3\over2}
\right]
\\ &&
-{N_cq^4\over6(4\pi)^2}
+{N_c\over3(4\pi)^2}
\left[q\sqrt{{q^2\over4}-\Delta^2}(26\Delta^2+q^2)
-12\Delta^2(\Delta^2+q^2)\log{{q\over2}+\sqrt{{q^2\over4}-\Delta^2}\over\Delta}
\right]
\theta(\mbox{$q\over2$}-\Delta)
\;,
\label{veff}
\eqa
\end{widetext}
where the argument $\Lambda$
indicates that the renormalized parameters are running
and the subscript $\ms$ indicates the scheme.
They satisfy the following renormalization group equations:
\bqa
\label{run1}
\Lambda{dm_{\ms}^2(\Lambda)\over d\Lambda}&=&{8 N_c m^2_{\ms}(\Lambda)g_{\ms}^2(\Lambda)
\over(4\pi)^2}\;,
\\
\Lambda{dg_{\ms}^2(\Lambda)\over d\Lambda}&=&{8 N_c g_{\ms}^4(\Lambda)\over(4\pi)^2}
\;,
\\
\Lambda{d\lambda_{\ms}(\Lambda)\over d\Lambda}&=&{16N_c\over(4\pi)^2}
\left[\lambda_{\ms}(\Lambda) g^2_{\ms}(\Lambda)-6g^4_{\ms}(\Lambda)
\right]\;,
\\
\Lambda{dh_{\ms}(\Lambda)\over d\Lambda}&=&{4 N_c g^2_{\ms}(\Lambda)
h_{\ms}(\Lambda)\over(4\pi)^2}
\label{run4}
\;.
\eqa
The solutions to Eqs. (\ref{run1})--(\ref{run4}) are
\bqa
\label{sol1}
m_{\ms}^2(\Lambda)&=&
{m_0^2\over1-{4g_0^2N_c\over(4\pi)^2}
\log{\Lambda^2\over \Lambda_0^2}
}\;,
\\
g_{\ms}^2(\Lambda)&=&
{g_0^2\over1-{4g_0^2N_c\over(4\pi)^2}
\log{\Lambda^2\over \Lambda_0^2}
}\;,
\\
\lambda_{\ms}(\Lambda)&=&{\lambda_0-{48g_0^4N_c\over(4\pi)^2}
\log{\Lambda^2\over \Lambda_0^2}
\over\left(1-{4g_0^2N_c\over(4\pi)^2}
\log{\Lambda^2\over \Lambda_0^2}
\right)^2}\;,
\label{sol3}
\\
h_{\ms}(\Lambda)&=&
{h_0\over1-{2g_0^2N_c\over(4\pi)^2}
\log{\Lambda^2\over \Lambda_0^2}
}\;,
\label{sol5}
\eqa
\newpage
The parameters $m_0^2$, $g_0^2$, $\lambda_0$ and $h_0$, are the 
values of the running parameters at the scale $\Lambda_0$, 
where we choose $\Lambda_0$ to satisfy
\bqa
\log{\Lambda_0^2\over m_q^2}+F(m_\pi^2)+m_\pi^2F^{\prime}(m_\pi^2)
&=&0\;.
\label{l0}
\eqa
$F(m_\pi^2)$ and $m_\pi^2F^{\prime}(m_\pi^2)$ vanish in the chiral limit which
implies that $\Lambda_0=m_q$.
We can now evaluate Eqs. (\ref{osm1})--(\ref{hhh}) at $\Lambda=\Lambda_0$ to find
$m_0^2$, $\lambda_0$, $g_0^2$, and $h_0$. Inserting Eqs. (\ref{sol1})--(\ref{sol5}) into Eq. (\ref{veff})
using the results for $m_0^2$, $\lambda_0$, $g_0^2$, and $h_0$, 
we obtain the final result Eq. (\ref{fullb}).
$ $\\ \\



\end{document}